\documentclass[10pt, conference, compsocconf]{IEEEtran}

\usepackage{amsmath}
\usepackage{graphicx,psfrag,epsf}
\usepackage{enumerate}
\usepackage{url} 
\usepackage{framed}
\usepackage{caption}
\usepackage[usenames, dvipsnames]{color}
\usepackage[table]{xcolor}

\begin{document}
%
\title{Inform Product Change through Experimentation with Data-Driven Behavioral Segmentation}

\author{\IEEEauthorblockN{Zhenyu Zhao, Yan He, Miao Chen}
\IEEEauthorblockA{Yahoo Inc.\\
Sunnyvale, USA \\ 
Email: miaoc@yahoo-inc.com}
}



\maketitle

\begin{abstract}
Online controlled experimentation is widely adopted for evaluating new features in the rapid development cycle for web products and mobile applications. Measurement on overall experiment sample is a common practice to quantify the overall treatment effect. In order to understand why the treatment effect occurs in a certain way, segmentation becomes a valuable approach to a finer analysis of experiment results. 
This paper introduces a framework for creating and utilizing user behavioral segments in online experimentation. By using the data of user engagement with individual product components as input, this method defines segments that are closely related to the features being evaluated in the product development cycle. With a real-world example we demonstrate that the analysis with such behavioral segments offered deep, actionable insights that successfully informed product decision making.
\end{abstract}

\section{Introduction} \label{intro}
The art of fast failure has been called the true secret of Silicon Valley's success \cite{Thrun2014}. Web product development is often composed of a build-measure-learn cycle \cite{ries2011lean}. During this cycle, new ideas, e.g., product changes in either front-end design or back-end algorithms, are evaluated and implemented at a fast pace. Online controlled experiments, also known as A/B tests, are widely used to evaluate these ideas \cite{Keppel1992, KOHAVI2007, KOHAVI2009, KOHAVI2013, KOHAVI2014, Tang2014recsys, Frasca2014recsys, Smallwood2014recsys, zhao2016online}. 

A standard experiment analysis measures the overall test effect by comparing all the test users with all the control users. The result of such analysis answers the ``what'' question - what will happen if the feature being tested is launched. However, it falls short of answering the ``why'' question - why the treatment effect occurs in a certain way. 

In order to answer the ``why'' question, segmentation is required to study the impact on individual user subgroups. Owing to the different intents and preferences of users, their reaction to product change can differ, which is often overlooked in experiment analyses involving only the overall population. Conducting experiment analysis at the user segment level will help the stakeholders understand 1) how the treatment effect varies among user subgroups, and 2) how much each subgroup contributes to the overall treatment effect. The answers to these questions are indispensable for guiding product development. 

It can be challenging to define user segments that are useful and relevant to experiment analyses. Common dimensions, such as device, browser, operating system, location, demographics, user tenure and referral source, are static and relatively easy to implement. In some cases these dimensions are useful for analysis and debugging. Additional dimensions that are directly related to the feature being evaluated in the experiment can add value to inform the development of products.

Behavioral segmentation using distance-based clustering and probability-based mixture models has been extensively studied in the fields including user targeting in marketing \cite{wedel2012market}, online advertising \cite{tu2010topic} \cite{yan2009much}, personalizing content serving \cite{hofgesang2007web}, identifying user behavior on social networks \cite{maia2008identifying}, and modeling shared interests of e-commerce users \cite{zhou2006web}. Applying segmentation in experimentation was also discussed in previous literature. Segmentation has been used as an experiment inclusion criteria for targeting subpopulations based on static and dynamic user attributes (\cite{xu2015infrastructure}). When data quality issues or unexpected results arise from an experiment, segmenting the experiment results (e.g. by browser type) can help identify the root cause (\cite{kohavi2011unexpected}). Segmentation was also used in a stratified sampling method for enhancing experiment accuracy  (\cite{xie2016improving}). 

In this paper, we present an approach to behavioral segmentation designed for experiment analysis. Users are clustered according to their level of engagement with individual product functional components. These mutually exclusive and collectively exhaustive segments are closely related to the features being evaluated and facilitate a finer analysis of experiment results.

The contribution of this paper includes:
\begin{itemize}
\item A novel framework is introduced for developing and utilizing behavioral segments tailored for experiment analysis. In particular, we design segmentation features that describe user intent and needs which are critical to guide the development of product features.

\item We demonstrate that the application of this method led to a successful product decision in a real-world example for Yahoo Finance.

\item Potential caveats and pitfalls are discussed when behavioral segmentation is used together with experimentation.

\end{itemize}

This paper is organized as follows: Section~\ref{sec:segmentation_model} presents the procedure for developing behavioral segmentation using a data-driven algorithm; Section~\ref{sec:segmentation_utilization} describes how to draw actionable insights through segment-level analyses in experimentation;
Section~\ref{sec:example} discusses a study with the proposed method applied in the Yahoo Finance website redesign experiment, which led to a product decision with significant improvement of user engagement. Section~\ref{sec:discussion} summarizes our research and discusses the potential pitfalls in practice.

\section{Developing Data-Driven Behavioral Segmentation for Online Experimentation}
\label{sec:segmentation_model}

\subsection{Behavioral Segmentation Features} 
The first step is to define a list of user behavior features to be used in the clustering algorithm. Various types of behavior dimensions are available to choose. As an example, users can be divided into high/medium/low engagement groups based on their visiting frequency. For evaluating product features, we found using the engagement with product functional components is particularly relevant. By using such dimensions we are able to measure the experiment effect by segments and guide product development with actionable insights. 

As an example, Yahoo Finance website is a product with multiple web pages providing different utilities to users. Users coming to Yahoo Finance are looking for financial news articles, stock quotes, stock portfolio performance, market data, communication with other users on message boards, or personal finance advice. User behavior and utility needs can be characterized by the page views on different functional web pages. Therefore, the behavioral features for segmentation consist of finance homepage page views, quotes page views, watchlist (or portfolio) page views, article page views, stock charts page views, message board page views, personal finance page views, and other page views. 

\subsection{Segment Definition Period} 

After defining the list of behavioral features, we collect the data for calculating such features. It is critical to define user segments based on the data from the pre-experiment period instead of the experiment period. This setup enables us to carry out a fair comparison and measure the treatment effect on each user segment. Otherwise, bias can be introduced as the treatment effect may shift user behavior and user segments in the treatment group.  
In the segment definition period, a data set containing user behavioral features is collected to define user segments before experiments start. In the experiment period, a data set containing experiment metrics is collected to measure treatment effects at both the overall level and the segment level. 
The length of the segment definition period may vary from one to multiple weeks.  

\subsection{Feature Transformation and Dimension Reduction} 
\label{sec:feature_transformation}
A series of feature engineering procedures, including data transformation and dimension reduction, are applied to prepare the user features as inputs for the segmentation algorithm. This process can be summarized in following six steps.

\begin{itemize}
\item Step 1: Feature extraction from raw data.

\begin{framed}
Suppose there are $n$ users and $p$ features; let
$x^{(1)}_{ij}$ denote the value of the $j$th feature for the $i$th user ($i=1,...,n$ and $j=1,...,p$), with the notation $^{(s)}$ indicating the features at Step $s$. 
\end{framed}

\item Step 2: Orthogonalization. 
Segmentation features in the original scale are highly correlated with the visiting frequency of the users. 
To reduce the dominant effect of visiting frequency, we adopted an orthogonal design with: 1) the number of visiting days per week, 2) the number of sessions per visiting day, and 3) module-specific engagement per session. 

\begin{framed}
Days visited per week:
$x^{(2)}_{i1} = x^{(1)}_{i1} / w$, where $x^{(1)}_{i1}$ is the number of days visited for user $i$ calculated in Step 1, and $w$ denotes the number of weeks during the segment definition period;

Sessions per visited day:
$x^{(2)}_{i2} = x^{(1)}_{i2} / x^{(1)}_{i1}$, where $x^{(1)}_{i2}$ is the number of sessions for user $i$ calculated in Step 1;

Per-session engagement features:
$x^{(2)}_{ij} = x^{(1)}_{ij} / x^{(1)}_{i2}$, with $j=3,...,p$, where $x^{(1)}_{ij}$ is the module-specific engagement feature calculated in Step 1.
\end{framed}

\item Step 3: Data cleaning. We remove outliers (users with extremely high values for page views, clicks, etc.) from the segmentation training data set (\cite{outlier2017}). 

\begin{framed}
$x^{(3)}_{ij}$ denotes the cleaned feature data,  with $j=1,...,p$ and $i=1,...,n'$, after removing $n-n'$ outliers and re-indexing the users.
\end{framed}

\item Step 4: Log transformation. Many clustering algorithms are based on the Euclidean distance in the feature space. Measuring on the original scale may not accurately represent the behavioral difference among the users. For example, users with one page view should be more distant from those with zero page views than from those with two page views. To accommodate this nonlinear relationship, we transform the original feature space into log scale: $x=\log(1+x)$ for each feature $x$.   
\begin{framed}
$x^{(4)}_{ij} = \log(1+ x^{(3)}_{ij})$ for $i=1,...,n'$ and $j=1,...,p$.
\end{framed}

\item Step 5: Normalization. In order to assign equal weight (in the loss function) to each feature in the segmentation algorithm, we normalize the log-transformed features to have mean $0$ and standard deviation $1$.
\begin{framed}
$$x^{(5)}_{ij} = \frac{x^{(4)}_{ij} - \frac{1}{n}\sum_{k=1}^n  x^{(4)}_{kj} }{\sqrt{\frac{1}{n-1} (x^{(4)}_{ij} - \frac{1}{n}\sum_{k=1}^n  x^{(4)}_{kj})^2 }}$$ for $i=1,...,n'$ and $j=1,...,p$.
\end{framed}

\item Step 6: Dimension reduction. To avoid the curse of dimensionality (\cite{keogh2011curse}) and reduce the correlation among features, principle component analysis (PCA, \cite{jolliffe2002principal}) is applied before clustering. Only the top principle components are included for model development, with a sufficient proportion of the variance from the original feature space explained.

\begin{framed}
$x^{(6)}_{ij}$ denotes the $j$th principle component score for user $i$, for $i=1,...,n'$ and $j=1,...,p'$ ($p'<=p$ as the first $p'$ principle components are selected).

Let $X^{(5)}$ denote the $n \times p$ feature matrix with element $x^{(5)}_{ij}$ at the $i$th row and $j$th column. The principle component score matrix can be calculated by 
$$X^{(6)} = X^{(5)}  W $$ 
where matrix $W$ is the $p \times p$ loading matrix whose columns are the eigenvectors of $X^{(5) T} X^{(5)}$. Further, $x^{(6)}_{ij}$ is the element of the matrix at the $i$th row and $j$th column.
\end{framed}

\end{itemize}

\subsection{Clustering Algorithm} 
\label{subsec:Clustering Algorithm}

Various clustering algorithms are available for developing the segments based on user behavioral features. We hereby summarize the criteria for choosing the optimal algorithm. 

\begin{itemize}
\item  Scalability. Computation requirements for large-scale web data-based segmentation can be intense. For this reason, some clustering algorithms are disqualified owing to computational inefficiency. For instance, the Agglomerative clustering algorithm (\cite{beeferman2000agglomerative}), as a bottom-up hierarchical approach, is not scalable for online experimentation because of the heavy computational load in pair-wise distance computations between users. 

\item  Goodness of fit.  Each model has either an explicit or implicit assumption of the underlying data. The goodness of fit between the actual data and the model assumption is an important criterion for model evaluation. For example, GMMs (Gaussian mixture models \cite{reynolds2015gaussian}) assume that the data of each segment follows a multi-variate Gaussian distribution in the feature space, and the quality of the segmentation depends on whether the data is close to a Gaussian shape. In addition, some geometry clustering models, such as DBSCAN (Density-based spatial clustering of applications with noise, \cite{ester1996density}), are useful when the clusters have a specific shape. However, DBSCAN assumes the densities are similar among different clusters such that an appropriate parameter can be specified to define neighborhood. In addition, DBSCAN may encounter difficulty for defining the neighborhood distance threshold when the dimension of feature space is high (curse of dimensionality). In contrast, k-means does not have a strong assumption of the underlying data structure, and can outperform other models when the data does not fit their assumptions. 

\item  Interpretation. In order to inform product decisions, the developed segments
must represent distinct user groups and bear clear business meaning. For example, a good segmentation algorithm is expected to create clusters of relatively similar sizes (instead of one dominant giant cluster with many tiny clusters), and distinguishable clusters (instead of two clusters with similar behavior patterns and business meaning). 

\end{itemize}

After evaluating multiple clustering algorithms (including the Agglomerative algorithm, DBSCAN, k-means, and GMM), we chose k-means for our products. The reasons are: 1) k-means is scalable for massive data sets; 2) k-means has weak assumptions of the data structure which allows us to apply this method in various use cases; 3) the segments created by k-means effectively represent our distinct user groups. We encourage the other practitioners to determine their best algorithm based on their own use cases.

However, the selection of the clustering algorithm is meant to be made on a case-by-case basis: k-means can be suboptimal to other algorithms in other applications and data settings.

The data input for k-means is composed of the principle component features defined in previous feature engineering steps. The output of the algorithm contains the centroids of each cluster in the feature space, and a cluster label for each user.

The number of clusters $K$ is a tuning parameter in the k-means algorithm. Choosing an optimal $K$ is a key issue in building the model (\cite{hamerly2003learning}). 
There are a few criteria for choosing this parameter value. 
For a given $K$, the objective of the K-means algorithm is to reduce the loss function defined as the sum of the squared distance from user points to cluster centroids. However, this loss function is not directly applicable for optimizing $K$, because it is a monotonic decreasing curve w.r.t. $K$ - the greater the number of clusters, the smaller the loss function. 
Instead, BIC (Bayesian information criterion, \cite{kass1995reference}) and Davies\-Bouldin index (\cite{davies1979cluster}) can be used to determine the optimal $K$. 
A detailed evaluation is discussed in Section~\ref{sec:example}. 

\subsection{Segment Name and Behavioral Representation} 
The user cluster label output from k-means is numerical, from $0$ to $K-1$, which does not bear a business meaning. Each segment can be further named after its unique user behavior pattern defined by per-user metrics described in Step $2$ in Section~\ref{sec:feature_transformation}. As an example, for Yahoo Homepage web portal site, one segment with high mail clicks and low article clicks can be named as 'Mail only', or vice versa, as 'Content only'. The resultant segment names now reflect user behavior patterns and their intent coming to the site. 

\section{Segmentation Analysis for Experimentation}
\label{sec:segmentation_utilization}
We have discussed the creation of the behavioral segments. This section will introduce a series of analyses using these segments in experimentation.    

\subsection{Preliminary Segment Analysis} 
In order to understand how important each segment means for the product, the percentage of overall engagement and/or revenue contributed by each segment can be quantified. Here we define experiment metrics that will be used for decisions. Examples include user count, page views, sessions, clicks, time spent, revenue, etc. For each of the experiment metrics, we calculate the total for both overall and segment users. Percentage contribution by each segment is calculated by dividing the total metric of segment users by that of overall users. These percentages are additive and sum to $100\%$ across segments. Note that some segments may have small size, but contribute to a relatively large portion of user engagement and/or revenue.

\subsection{Segment-Level Treatment Effect} 
For a finer view of experiment analysis, segment-level treatment effect is calculated as the mean difference between the test group and the control group. 
Given that the segment definition period is separate from the experiment period, we first define three types of user status:
\begin{enumerate}
\item Users appearing in both the segment definition period and the experiment period. These users have segment label assigned in segment definition period, as well as experiment metrics measured in experiment period.
\item Users appearing only in the segment definition period and not in the experiment period. These users are assigned to a segment but do not have an experiment metric value due to inactivity. Their experiment metric value is assigned as $0$, indicating no engagement. 
\item Users only appearing in the experiment period, but not in the segment definition period. These users are either new or returning users who were not active during the segment definition period. They don't have segment label assigned, but do have experiment measurement. 
This group is named as segment ``unseen'' and added to the original segment family for further analysis.
\end{enumerate}

For segment ``unseen'' and each segment defined in Section \ref{sec:segmentation_model}, both the metric difference and statistical significance (in the form of a p-value) are calculated to quantify segment-level treatment effect. It is not rare to observe that segments react in opposite direction to the feature being tested. Segment-level analysis on treatment effect uncovers such information buried in overall measurement. It also offers unique insights into the opportunities where product experience should be improved for specific groups of users.

\subsection{Identifying the Dominant Segments for the Overall Treatment Effect}
This step helps identify the drivers of overall treatment effect if there is one or a few dominant segments. The segments with a large treatment effect on per-user metrics are not necessarily the ones that dominate the overall treatment effect, because segment size is another factor. 
In order to evaluate a segment's impact on the overall metrics, we first calculate the within-segment total metric difference between the control and test groups. Then, the percentage of overall treatment effect explained by one segment is calculated by dividing its within-segment total metric difference by the total metric difference in the overall experiment population. 
Segments that contribute to a high percentage of the overall treatment effect are important ones that worth special attention in the iteration of product testing.

\section{Example}
\label{sec:example}

Next we will illustrate how to apply the proposed methodology in practice. The entire process from developing behavioral segments to the findings and decisions based on the segmentation analysis is discussed in the setting of a website redesign test for Yahoo Finance site. Empirical challenges, findings and evaluation of the proposed methodology are also discussed with this example. The data described in this example has been modified and rescaled.

\subsection{Experiment Background} \label{sec:expbackground}
In Yahoo Finance redesign experiment, changes were applied to overall page layout, module positions, content density and addition/removal of functional modules. A functional module can be a navigational module with links to other pages, an article module containing an article stream, or a quotes module showing stock prices. The impact on user engagement was evaluated via experimentation. 

Measurement includes both page engagement and user visitation frequency. Visitation frequency is quantified by days visited and sessions in the defined experiment period. Page engagement is quantified by the number of page views. There are two sub-categories of page views defined at Yahoo: (1) CPV (classic page view) is defined as a page view event logged when the user open the web page; (2) APV (additive page view) is defined as a different type of page view event logged when the user scrolls down the web page for a certain number of pixels, such that the user is viewing different content compared with the initial screen when the page was initiated. Therefore, if there is an APV logged on a page, there must be a preceding CPV. Although these two metrics are often correlated, CPV measures how frequently a user visits a certain page, while APV measures the depth of content consumption on the page. 

The launch criteria for the new site was not met due to the decline observed in the initial experiment results. There was a significant decline by $10.3\%$ in CPV, a significant increase by $14.0\%$ in APV, and insignificant decline in days visited and sessions. The trade-off of CPV to increased APV was expected with the change of page layout. However, the magnitude of CPV decline was too large to neglect. Further analysis at segment level is required to pinpoint the areas for improvement in future iteration.  

\subsection{Segment Definition Period and Experiment Period} 
Before the formal experiment (A/B test) started, we set up a so-called A/A test \cite{KOHAVI2009}. In the A/A test, users were split into a control group and a test group without any change applied. A/A test enabled us to validate if there was any pre-existing difference between the test group and the control group. In this example, two weeks of A/A testing was performed, and the A/A test period was used as the segment definition period. A/B test started after the completion of A/A testing. One week of A/B test data were used for the experiment analysis. 

\subsection{User Features for Creating Segments} 

As described in Section \ref{sec:expbackground}, page engagement is measured by CPV and APV which are used as segmentation features. In addition, frequency metrics days visited per week and sessions per visited day, and search clicks are also included for defining user segments in this example. Below is the list of segmentation features measured for each user:

\begin{itemize}
\item \texttt{vdays\_per\_week}: Average number of visiting days per week on Yahoo Finance site. 
\item \texttt{sessions\_per\_vdays}: Average sessions per days visited.  
\item \texttt{articles\_cpv\_per\_session}: Average CPVs on article page per session. 
\item \texttt{articles\_apv\_per\_session}: Average APVs on article page per session.  
\item \texttt{charts\_cpv\_per\_session}: Average CPVs on stock charts page per session. 
\item \texttt{charts\_apv\_per\_session}: Average APVs on stock charts page per session.  
\item \texttt{homepage\_cpv\_per\_session}: Average CPVs on  Yahoo Finance homepage per session. 
\item \texttt{homepage\_apv\_per\_session}: Average APVs on  Yahoo Finance homepage per session.
\item \texttt{lookup\_cpv\_per\_session}:  Average CPVs on stock ticker symbol lookup page per session.
\item \texttt{lookup\_apv\_per\_session}: Average APVs on stock ticker symbol lookup page per session.
\item \texttt{market\_data\_cpv\_per\_session}: Average CPVs on market data page per session.  
\item \texttt{market\_data\_apv\_per\_session}: Average APVs on market data page per session.
\item \texttt{message\_boards\_cpv\_per\_session}: Average CPVs on message boards (user forum) per session. 
\item \texttt{message\_boards\_apv\_per\_session}: Average APVs on message boards per session.  
\item \texttt{personal\_finance\_cpv\_per\_session}: Average CPVs on personal finance page per session.
\item \texttt{personal\_finance\_apv\_per\_session}: Average APVs on personal finance page per session.
\item \texttt{quotes\_cpv\_per\_session}: Average CPVs on individual stock quote page per session. 
\item \texttt{quotes\_apv\_per\_session}: Average APVs on individual stock quote page per session.
\item \texttt{watchlist\_cpv\_per\_session}: Average CPVs on customized stock watch list page per session. 
\item \texttt{watchlist\_apv\_per\_session}: Average APVs on customized stock watch list page per session.
\item \texttt{other\_cpv\_per\_session}: Other CPVs per session. Other CPVs are defined as CPVs that do not belong to the pages in the categories above. 
\item \texttt{other\_apv\_per\_session}: Other APVs.
\item \texttt{search\_clicks\_per\_session}: Number of clicks on the search button per session. 
\end{itemize}

The feature engineering process described in Section~\ref{sec:feature_transformation} was applied to the listed segmentation features. Out of the $23$ user features, $14$ principle components were selected that explained $85\%$ of the variance in the original feature space. 

\subsection{Creating Segments Using K-means} 
\begin{figure}[ht]
\includegraphics[scale=0.42]{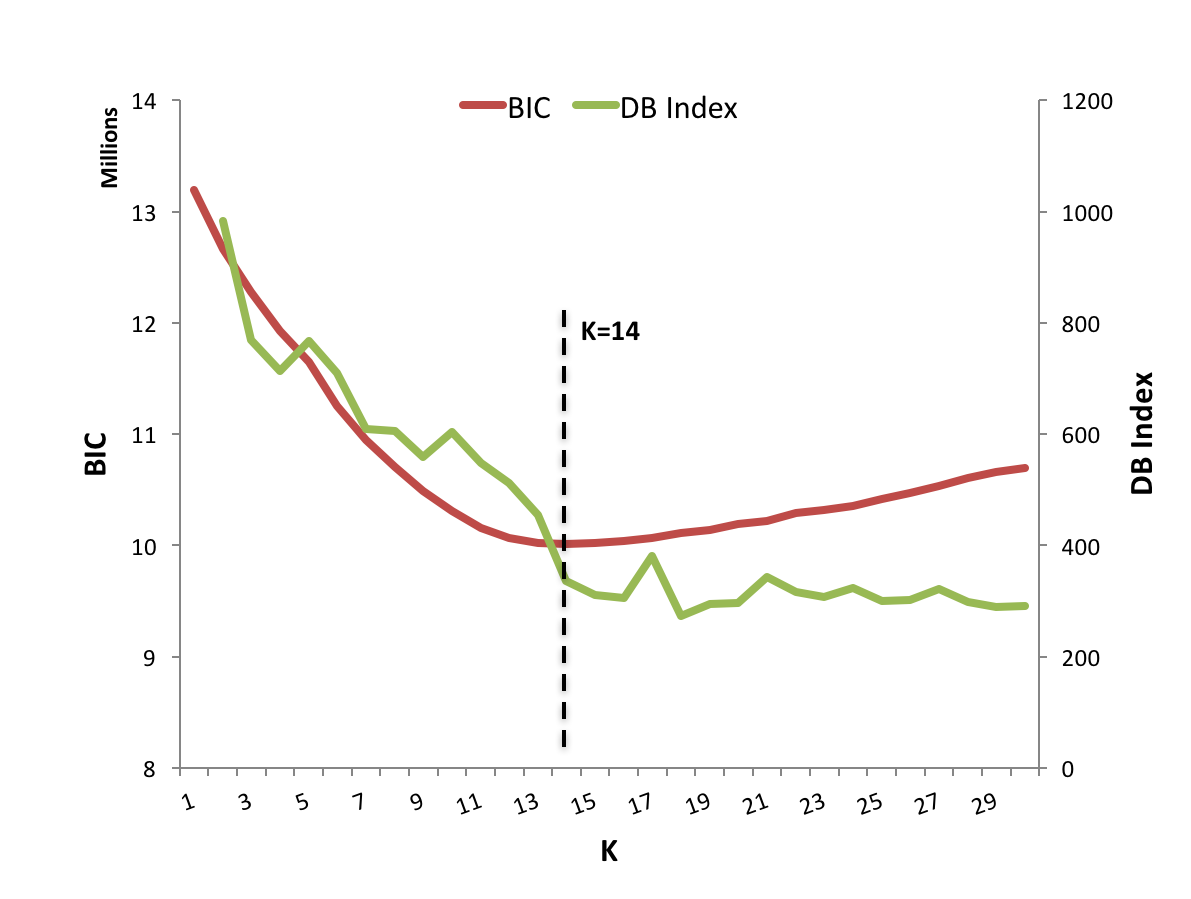}
\caption{BIC and Davies-Bouldin Index for k-means with different number of clusters K}
\label{fig:kmeans_loss_bic_dbindex}
\end{figure}

\begin{figure*}[ht]
\begin{center}
\includegraphics[scale=0.85]{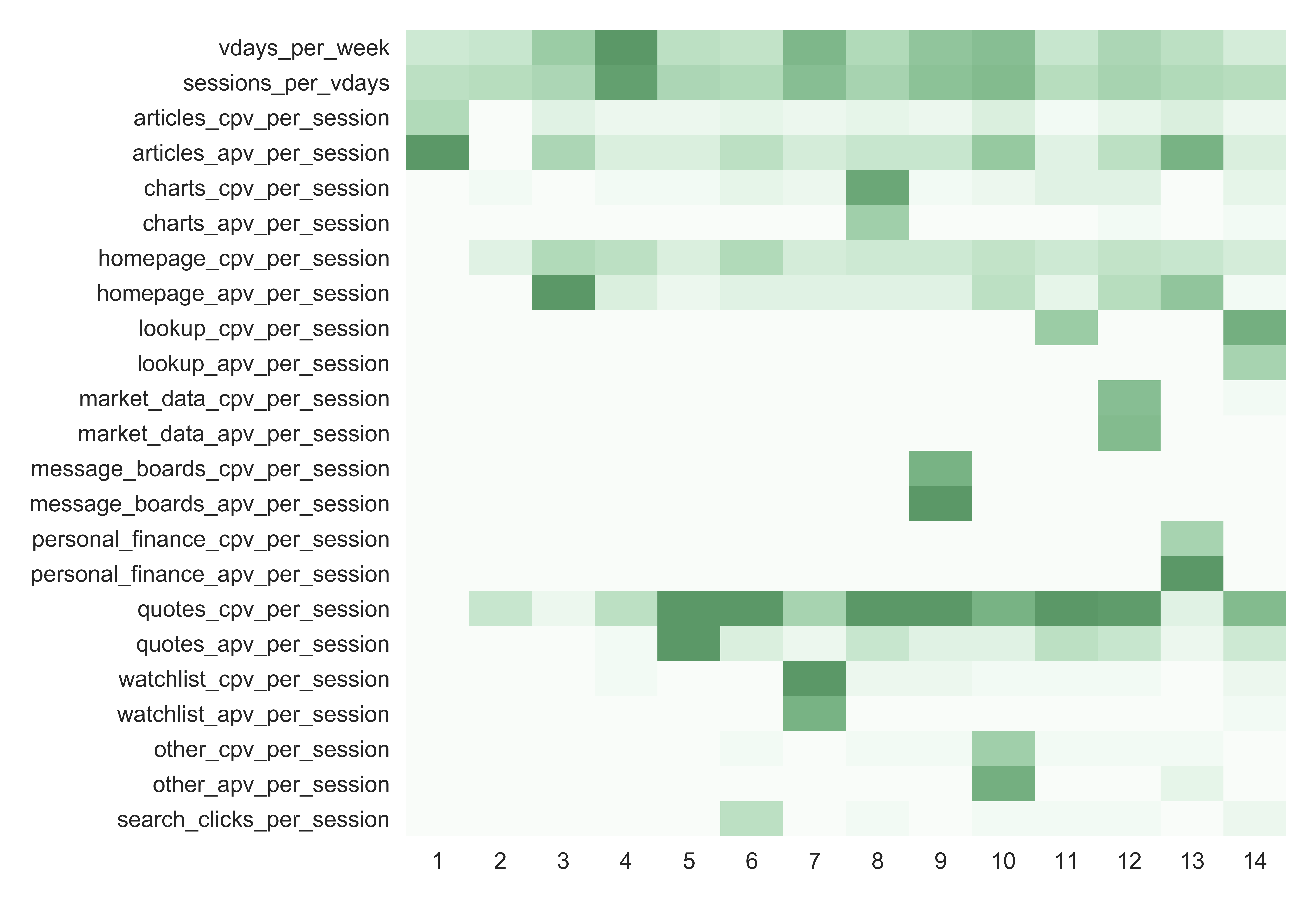}
\caption{Average Per-User Metrics for Each Cluster Defined by K-means Algorithm for Yahoo Finance. Green indicates high engagement, while white denotes low engagement.  
There are a total of $14$ clusters defined by the k-means algorithm. The x-axis shows the cluster label from k-means.}
\label{fig:seg_def2}
\end{center}
\end{figure*}

A k-means clustering algorithm was applied on the principle component feature space to create the user clusters. As discussed, parameter $K$ must be chosen based on certain criteria. Different $K$ values ($1,2,3,...,30$) are evaluated by the BIC and Davies\-Bouldin index in Figure ~\ref{fig:kmeans_loss_bic_dbindex}. 
When $K=14$, both loss functions are close to the minimum; therefore, $K=14$ was chosen in this case study. The remaining paper would assume $14$ clusters are adopted. Figure ~\ref{fig:seg_def2} shows $14$ clusters defined by k-means, and the average per-user metrics measured in each cluster. 

\subsection{Behavioral Segment Name}

Segments are named after the user engagement pattern defined in Figure ~\ref{fig:seg_def2}. Note that each segment is evaluated at different features in original scale, instead of in the principle component space. For example, segment $1$  is named as ``Article Only'' because it has a high number of article page views and low engagement on other pages. Segment $2$ is named as ``Tourist \& Quotes'' with a relatively high quotes page CPV and generally low engagement everywhere. Table ~\ref{tab:seg_name} shows all the behavioral segment names inferred from the user engagement pattern. 

\begin {table}
\caption {Behavioral Segment Names for k-means Clusters}
\label{tab:seg_name}
{\small
\begin{tabular}{ | l | l | }
\hline
K-means Cluster &	Behavioral Segment Name\\
\hline
Cluster 1  & Article Only  \\ 
Cluster 2  & Tourists \& Quotes  \\ 
Cluster 3 & Homepage \& Hybrid Moderate  \\ 
Cluster 4 & Homepage \& Hybrid High  \\ 
Cluster 5 & Quotes Only  \\ 
Cluster 6 & Quotes \& Search  \\ 
Cluster 7 & Watchlist Only  \\ 
Cluster 8 & Quotes \& Charts  \\ 
Cluster 9 & Quotes \& Message Board  \\ 
Cluster 10 & Quotes \& Other  \\ 
Cluster 11 & Quotes \& Lookup  \\ 
Cluster 12 & Quotes \& Market Data  \\ 
Cluster 13 & Personal Finance  \\ 
Cluster 14 & Lookup \& Quotes  \\ 
\hline
\end{tabular}
}
\end{table}

\subsection{Preliminary Segment Analysis}

\begin{figure}[ht]
\includegraphics[scale=0.3]{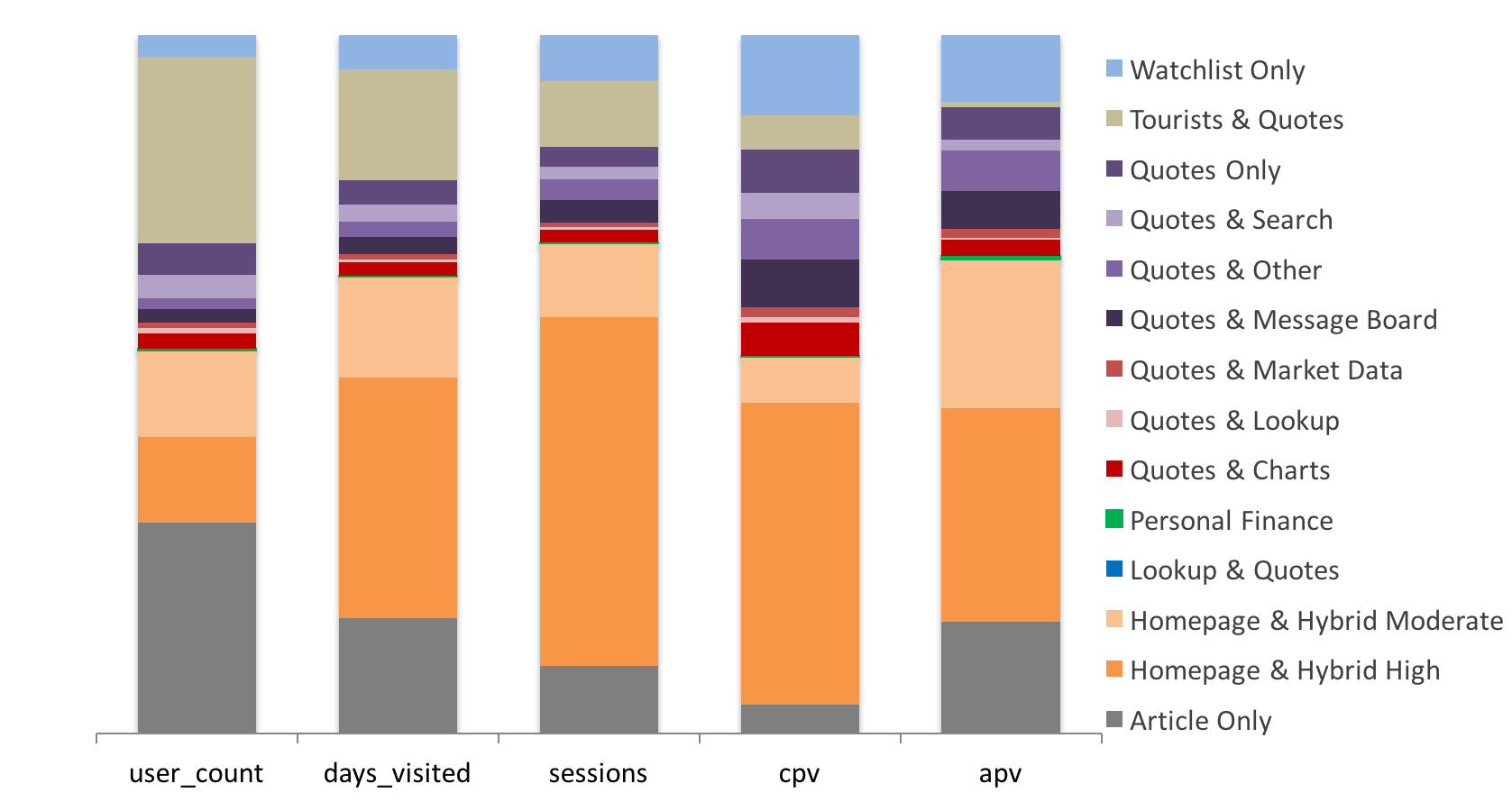}
\caption{Preliminary View of Segment Importance Measured by Contribution to Experiment Metrics}
\label{fig:aa_perc}
\end{figure}

Figure ~\ref{fig:aa_perc} shows the percentage of each segment's contribution to four total engagement metrics (user\_count, days\_visited, sessions, and CPV) during the segment definition period. It is worth noting that the segments' contribution to user engagement is not proportional to segment size (user count). 
Some larger segments including ``Tourist \& Quotes'' and ``Article Only'' contribute to a relatively small proportion of user engagement. Conversely, smaller segments such as ``Homepage \& Hybrid High'', ``Watchlist Only'', ``Quotes \& Message Board'' and ``Quotes \& Other'' contribute to a large proportion of user engagement relative to their size. 
This analysis revealed that ``Homepage \& Hybrid High'' is the main contributor to overall engagement (days\_visited, sessions, and CPV).

\subsection{Segment-Level Treatment Effect} \label{seg-effect}

\begin {table*}
\begin{center}
\caption {Segment Treatment Effect: Difference in Average Per-User Metrics between Test and Control by Segment (n/a is shown for segments whose sample sizes are too small for analysis ($<1\%$))}
\label{tab:ab-seg-per-user}
{\small
\begin{tabular}{ | l | r | r | r | r | }
\hline
Segment & Users & days\_visited  & sessions & cpv \\
\hline
Total & $100\%$ & $-1\%$ & $-2\%$ & $-10\%$ \\
\hline
Article Only & $10\%$ & $0\%$ & $0\%$ & $3\%$ \\
\cellcolor{yellow}Homepage \& Hybrid High & $15\%$ & $-2\%$ & $-5\%$ & $-15\%$ \\
Homepage \& Hybrid Moderate & $7\%$ & $-3\%$ & $-3\%$ & $-8\%$ \\
Lookup \& Quotes & $< 1\%$ & n/a & n/a & n/a  \\
Personal Finance & $< 1\%$ & n/a & n/a & n/a  \\
\cellcolor{yellow}Quotes \& Charts & $1\%$ & $-6\%$ & $-9\%$ & $-12\%$  \\
Quotes \& Lookup & $< 1\%$ & n/a & n/a & n/a  \\
Quotes \& Market Data & $< 1\%$ & n/a & n/a & n/a   \\
\cellcolor{yellow}Quotes \& Message Board & $1\%$ & $-3\%$ & $-6\%$ & $-19\%$ \\
Quotes \& Other & $1\%$ & $2\%$ & $-1\%$ & $-20\%$ \\
\cellcolor{yellow}Quotes \& Search & $1\%$ & $-5\%$ & $-10\%$ & $-26\%$ \\
Quotes Only & $2\%$ & $0\%$ & $0\%$ & $-10\%$ \\
Tourists \& Quotes & $9\%$ & $-1\%$ & $2\%$ & $4\%$ \\
\cellcolor{yellow}Watchlist Only & $2\%$ & $-2\%$ & $-8\%$ & $-9\%$ \\
Unseen & $49\%$ & $-1\%$ & $1\%$ & $-5\%$ \\
\hline
\end{tabular}
}
\end{center}
\end{table*}

During the experiment period, the treatment effect in each segment can be measured by comparing the control and test users within the segment. In Table ~\ref{tab:ab-seg-per-user}, the treatment effect is measured by three per-user metrics: days\_visited, sessions and CPV. Segments ``Homepage \& Hybrid High'', ``Quotes \& Charts'', ``Quotes \& Message Board'', ``Quotes \& Search'', and ``Watchlist Only'' have higher-than-average declines comparing to the overall population (``Total'').
These segments are highlighted in yellow. In contrast, other segments have smaller decline or even an increase in user engagement, such as ``Article Only'', ``Tourists \& Quotes'' and ``Unseen''. Note that segments ``Lookup \& Quotes'', ``Personal Finance'', ``Quotes \& Lookup'' and ``Quotes \& Market Data'' have extremely small sample sizes. As a result, the results of these segments are labeled as ``n/a''. This analysis indicated negative user experiences on Homepage, Quotes, and Watchlist should be addressed and corrected. 

\subsection{Identifying High-impact Segments} \label{domsegs}

\begin{figure}[ht]
\begin{center}
\includegraphics[scale=0.3]{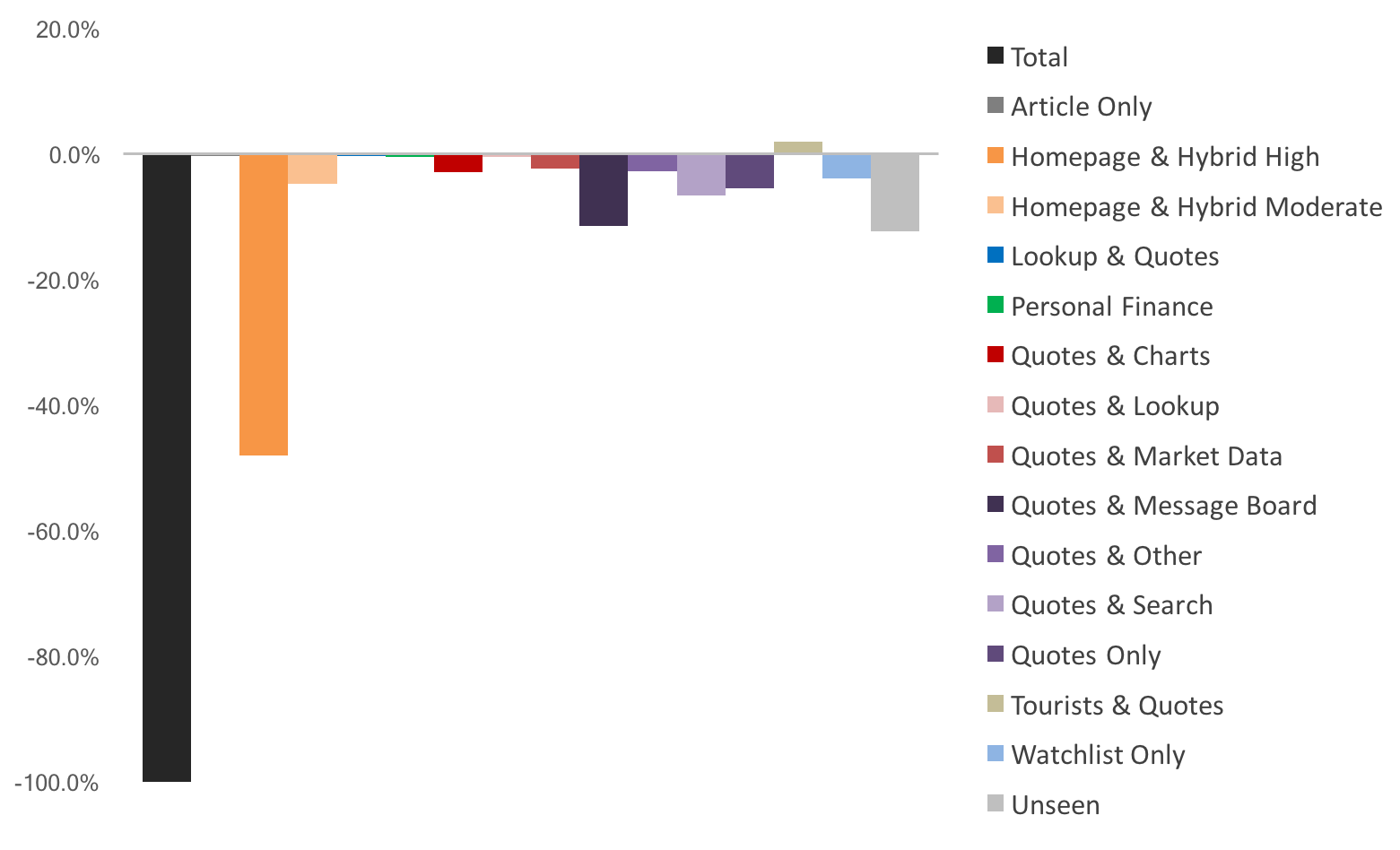}
\caption{Percentage Contribution to Total Drop in CPV by Segment}
\label{fig:ab-seg-agg}
\end{center}
\end{figure}

In this step we will determine the driver segments of the overall engagement decline, particularly in CPV. 
Figure ~\ref{fig:ab-seg-agg} shows the proportion of each segment's contribution to the overall CPV decline. The top three high-impact segments are: ``Homepage \& Hybrid High'' ``Quotes \& Message Board'' and ``Unseen''. The ``Unseen'' segment has a large sample size but its average per-user decline is relatively small as shown in Table ~\ref{tab:ab-seg-per-user}.  Based on the results from this step and Section \ref{seg-effect}, the team decided to identify and fix the user experience issues for segments ``Homepage \& Hybrid High'' and ``Quotes \& Message Board'' as a priority.

\subsection{Follow-up Root Cause Analysis} 

\begin{figure}[ht]
\begin{center}
\includegraphics[scale=0.3]{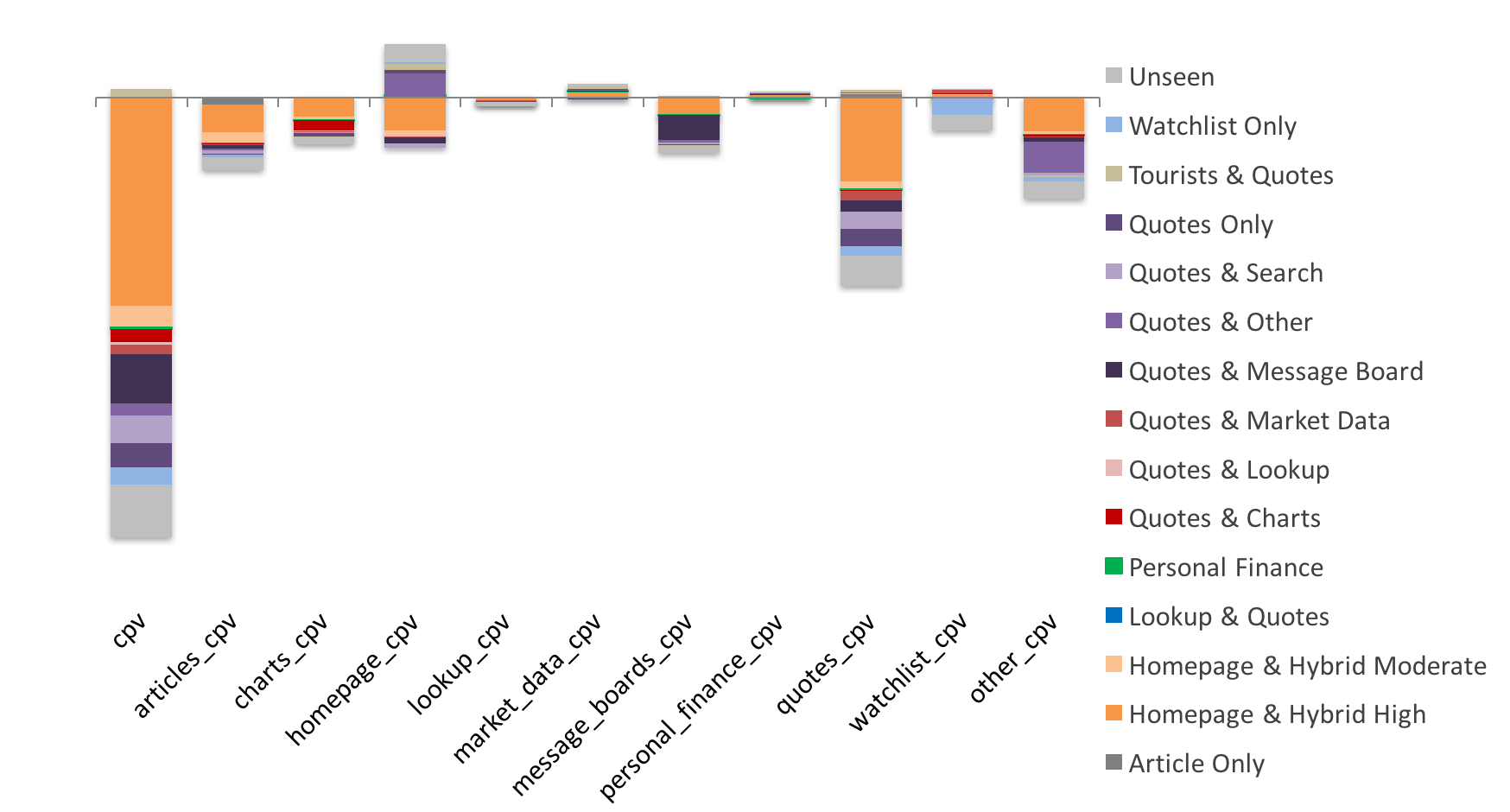}
\caption{CPV Difference between Test and Control by Page Type and Segment}
\label{fig:ab-seg-page-type}
\end{center}
\end{figure}

In order to pinpoint the root cause of CPV drop for the two high-impact segments identified in \ref{domsegs}, we further measured CPV by page types. Figure ~\ref{fig:ab-seg-page-type} shows that segment ``Homepage \& Hybrid High'' lost most CPV on quotes page. Interestingly, the CPV decline on finance homepage itself was smaller for this homepage segment. Because finance homepage is a main source of traffic referral to quotes page, we suspected that path from finance homepage to finance quotes page was blocked. Comparing the interface of new design to the old, we determined that this drop was due to the removal of two quotes-related modules. Original motivation of removing these two modules was to create space for new features; however, the large drop in quotes CPV was critical and should be fixed.

\begin {table*}
\begin{center}
\caption {Overall Treatment Effect (\%  Difference is defined as the relative difference (using the control mean as a baseline) between the treatment mean and control mean of the per-user level metric)}
\label{tab:after-fix-ab-overall-per-user}
{\small
\begin{tabular}{ | l | r | r | r | r | }
\hline
& \multicolumn{2}{c|}{Before Fix} & \multicolumn{2}{c|}{After Fix} \\
\hline
Metric & \%  Difference & Significant & \%  Difference & Significance \\
\hline
days\_visited & $-0.7\%$  & Not Significant ($p > 0.05$) & $0.3\%$  & Not Significant ($p > 0.05$) \\
sessions & $-1.9\%$  & Not Significant ($p > 0.05$) & $-1.2\%$  & Not Significant ($p > 0.05$) \\
cpv & $-10.3\%$  & Significant ($p < 0.05$) & $-6.1\%$  & Significant ($p < 0.05$) \\
apv & $14.0\%$  & Significant ($p < 0.05$) & $35.1\%$  & Significant ($p < 0.05$) \\
\hline
\end{tabular}
}
\end{center}
\end{table*}

\begin {table}
\begin{center}
\caption {Segment Level Treatment Effect (\%  Difference is defined as the relative difference (using the control mean as a baseline) between the treatment mean and control mean of the per-user level metric)}
\label{tab:after-fix-ab-seg-per-user}
{\small
\begin{tabular}{ | l | r | r | }
\hline
& \multicolumn{2}{c|}{\% Difference in cpv}  \\
\hline
Segment & Before Fix & After Fix \\
\hline
Homepage \& Hybrid High & $-15\%$ & $-9\%$ \\
Homepage \& Hybrid Moderate	& $-8\%$ & $6\%$ \\
\hline
\end{tabular}
}
\end{center}
\end{table}

For segment ``Quotes \& Message Board'', the decline was driven by the CPV loss in message board as shown in Figure ~\ref{fig:ab-seg-page-type}. There are separate message board forums for test group and control group. Test group has a much smaller sample size than control group by design, as a result, there are much less user-generated content on the test message board compared to the control message board. Therefore, the CPV drop observed in segment ``Quotes \& Message Board'' can be explained by this network effect. This decline is expected and should recover when the new design is released to $100\%$ population.

\subsection{Impact on Product Development} 
Based on the analysis described above, we added back two quote-related modules on finance homepage. 
After the fix, another round of experiment was performed to verify the result. Table ~\ref{tab:after-fix-ab-overall-per-user} showed that the fix resulted in a sizable recovery of CPV as well as a larger increase in APV. Table ~\ref{tab:after-fix-ab-seg-per-user} further verified that significant improvement happened to homepage-related segments ``Homepage \& Hybrid High'' and ``Homepage \& Hybrid Moderate''.

\section{Conclusions}
\label{sec:discussion}

Controlled experimentation has become an integral part of iterative development cycle of web products and mobile applications \cite{ries2011lean}. In this paper we described how to define and utilize user behavioral segments in experimentation. By using the data of user engagement with individual product components as input, this method defines segments that are closely related to the features being evaluated in an A/B test. Analysis with these segments offers deep, actionable insights by answering the ``why'' question raised in Section \ref{intro}. In comparison to a direct page view analysis by paths or referrals, segmentation enables experiment owners to study a broader range of metrics including frequency metrics such as sessions. In the example of Yahoo Finance site redesign experiment, we demonstrated that the proposed method successfully improved user engagement by informing a critical design fix on Finance homepage. 

When segmentation is used together with experimentation, several pitfalls have been discussed in previous works. For example, running an analysis on a large number of segments may lead to inflated false positives (\cite{KOHAVI2013}). Simpson’s paradox may arise when combining segment results (\cite{crook2009seven}). Further, the experiment effect on one segment cannot be simply generalized to the overall population (\cite{KOHAVI2014}). 

It is important to note that the defined segments must be independent of the experiment treatment. Otherwise, it can introduce bias in the analysis. For example, if user segments are created based on their behavior during the experiment period, comparison between the treatment group and the control group will become invalid at segment level. This is because the behavior pattern of treatment group may be shifted in the experiment period. In the framework proposed in this paper, this bias is prevented by defining user segments using pre-experiment data which is not impacted by the treatment effect. 

In addition to the provided example, we have applied the same approach in various product tests at Yahoo. In future research, we look to develop a scalable, computationally economical solution to the integration of behavioral segmentation with our experimentation platform.

\section*{Acknowledgment}
We would like to extend our thanks to Maria Stone and Jon Malkin for insightful discussion on this study, to Don Matheson, Nikhil Mishra and Abhinav Chaturvedi for implementing our work in an analytics pipeline, and to Prashant Ramarao, Charles Hartel and Trang Le for setting up the Yahoo Finance experiment and developing the redesigned product to validate the fix.

\bibliographystyle{./IEEEtran}
\bibliography{./IEEEabrv,./IEEEexample}

\end{document}